\documentclass[runningheads,a4paper]{llncs}
\usepackage{makeidx}  
\usepackage{array}
\usepackage{url}
\usepackage{hyperref}
\usepackage{graphicx}
\usepackage{listings}
\usepackage{booktabs} 
\usepackage{longtable}
\urldef{\mailsa}\path|{david.kelk@uoit.ca, davidthomas.devine@gmail.com}|    
\newcommand{\keywords}[1]{\par\addvspace\baselineskip
\noindent\keywordname\enspace\ignorespaces#1}

\begin{document}

\mainmatter

\title{A Scienceographic Comparison of Physics Papers from the arXiv and viXra Archives}

\titlerunning{Scienceographic Comparison of arXiv, viXra}

\author{David Kelk$^{1}$ \and David Devine}

\authorrunning{David Kelk, David Devine}

\institute{$^{1}$University of Ontario Institute of Technology\\Oshawa, ON, Canada\\
\url{http://www.uoit.ca}\\
\mailsa\\}

\toctitle{A Scienceographic Comparison of Physics Papers From the arXiv and viXra Archives}
\tocauthor{David Kelk and David Devine}
\maketitle

\begin{abstract}
arXiv is an e-print repository of papers in physics, computer science, and biology, amongst others. viXra is a newer repository of e-prints on similar topics. Scienceography is the study of the writing of science. In this work we perform a scienceographic comparison of a selection of papers from the physics section of each archive. We provide the first study of the viXra archive and describe key differences on how science is written by these communities.
\end{abstract}

\section{Introduction}\label{sec:Introduction}

\textit{Bibliometrics}~\cite{DB09,BF02} is the application of mathematical and statistical methods to books and other media of communication~\cite{HW01}. \textit{Citation analysis}~\cite{DB09,M05} and \textit{content analysis}~\cite{K04} are two bibliometric methods.  \textit{Scientometrics}~\cite{LM12,HW01} is bibliometrics applied to science and technology. Both bibliometrics and scientometrics are applied after publishing.  \textit{Scienceography}~\cite{CMY12} is the study of the writing of science. How papers are written and how they describe their results is an under-studied area. This paper performs a scienceographic analysis of 20 papers from the physics categories of the arXiv and viXra repositories. (40 papers total.) We examine whether or not there are differences in how these communities of authors write science. We were especially interested in learning if there are metrics indicating that a paper is from one archive or the other.

arXiv\footnote{arxiv.org} is a repository and distribution server for research papers.  It was started in August, 1991 and hosts approximately 771,000 documents\footnote{Retrieved on July 19, 2012}. Part of its \textit{Goals and Mission} statement reads:

\begin{quote}
arXiv is an openly accessible, moderated repository for scholarly articles in specific scientific disciplines. Material submitted to arXiv is expected to be of interest, relevance, and value to those disciplines. arXiv reserves the right to reject or reclassify any submission. Submissions are reviewed by expert moderators to verify that they are topical and refereeable scientific contributions that follow accepted standards of scholarly communication (as exemplified by conventional journal articles)\footnote{arxiv.org/help/primer, Retrieved July 19, 2012}.   
\end{quote}

viXra\footnote{vixra.org} (arXiv spelled backwards) is a more recent e-print archive.  It was started in July, 2009 and hosts approximately 3,140 papers\footnote{Retrieved on July 19, 2012}.  viXra was created as a reaction to arXiv:

\begin{quote}
ViXra.org is an e-print archive set up as an alternative to the popular arXiv.org service owned by Cornell University. It has been founded by scientists who find they are unable to submit their articles to arXiv.org because of Cornell University's policy of endorsements and moderation designed to filter out e-prints that they consider inappropriate.
\end{quote}
\begin{quote}
ViXra is an open repository for new scientific articles. It does not endorse e-prints accepted on its website, neither does it review them against criteria such as correctness or author's credentials.\footnote{vixra.org, retrieved July 19, 2012} \footnote{There is no connection or affiliation between the arXiv and viXra sites.  We take no position on the archives, their policies, or the papers reviewed here.}
\end{quote}

In total, the 40 papers examined\footnote{Raw data is available \href{https://docs.google.com/spreadsheet/ccc?key=0AhxoA9AVV1G9dFNralJZUm5RMF9Ta195bGRINTVXakE}{here} (Google drive)} had 486 pages, 60 authors, 1040 numbered equations and 751 references. We found there were differences between the papers in each archive. arXiv papers had on average 1.9 authors who were always affiliated with a university or an equivalent institution. viXra papers had an average of 1.1 authors who were university affiliated 35\% of the time.  65\% of arXiv papers were published in journals while 55\% of viXra papers were published as web pages. arXiv papers averaged 14.3 pages in length, had a total of 481 numbered equations and 521 references. viXra papers averaged 9.9 pages, had a total of 559 numbered equations and 230 references. The complete data set is listed in Tables~\ref{tbl:Area1} through~\ref{tbl:Area4}.

We found there are indicators identifying the source archive. If some or all of the authors had a university or equivalent institution affiliation, or the paper contained theorems or lemmas, it was likely to be from the arXiv repository. If a paper was published in web pages, had inline citations not in the references section, or had web hyperlinks outside the references section it was likely to be from the viXra archive. (See Table~\ref{tbl:RQ1}.)

The rest of this paper is organized as follows. Our survey methodology and research question is introduced in section~\ref{sec:SurveyMethodology}. Results are described in section~\ref{sec:SurveyResults}. Related work is discussed in~\ref{sec:RelatedWork} and is followed by threats to validity in~\ref{sec:ThreatsToValidity}. Conclusions and future work are discussed in section~\ref{sec:Conclusions}.

\section{Survey Methodology}\label{sec:SurveyMethodology}

Both the arXiv and viXra e-print servers are divided into a number of top-level categories: physics, mathematics, computer science and others. This paper studied only the physics category.  Papers from the years 2007-2012 were considered from arXiv and 2009-2012 for viXra, due to the latter's more recent creation. Twenty papers were selected from each archive.  Papers were chosen in pairs: a paper was chosen from the same sub-category from each site. As the sub-category names don't match up exactly we chose the closest matching one. For example, if we chose a paper from \textit{High Energy Physics - Phenomenology} from arXiv, a paper from \textit{High Energy Particle Physics} was chosen from viXra. Year and month of submission for each member of the pair were chosen randomly. To reduce bias, the paper at numerical position 10 was always selected. For viXra there were a number of months where the total number of submissions was under 10. In this case we randomly chose a year and month with 10 or more submissions. If no month had 10 or more submissions, we always selected the paper at numerical position 5. This latter case didn't occur in the survey.

Data collection was divided into 4 broad areas and guided by our research question, summarized in Table~\ref{tbl:Approach}.

\begin{table}[t!]
\centering
\begin{tabular}{m{2cm}m{10cm}}
\hline
\textbf{Item} &
\textbf{Description}
\\\hline

Area 1 & Authors, their affiliation and collaboration  
\\

Area 2 & Publishing and citation
\\

Area 3 & Writing metrics (\# versions, \# pages, \# theorems and lemmas, \ldots)
\\

Area 4 & Referencing
\\

RQ 1 & Do any of the metrics identify a paper as coming from arXiv or viXra?
\\\hline
\end{tabular}
\vspace{2mm}
\caption{Research areas and research question.}
\label{tbl:Approach}
\end{table}

\section{Survey Results}\label{sec:SurveyResults}

Each of the four survey areas are examined in subsections~\ref{sec:AAC} to~\ref{sec:Ref} and summarized in Tables~\ref{tbl:Area1} to~\ref{tbl:Area4}.  Our research question is answered in subsection~\ref{sec:MISA} and Table~\ref{tbl:RQ1}.

\subsection{Authors, Affiliation and Collaboration}\label{sec:AAC}

More than half of the selected arXiv papers had two or more authors, all of whom were affiliated with a university or equivalent institution.  There was a moderate amount of collaboration between institutions\footnote{Authors from the same institution do not count as collaboration for this metric.}. (See Table~\ref{tbl:Area1}.) In contrast, the vast majority of viXra papers surveyed had one author who was not likely to be affiliated with a university.  It also follows then the rate of collaboration was very low.  


\begin{table}[t!]
\centering
\begin{tabular}{m{5cm}m{2cm}m{2cm}}
\hline
\textbf{Metric} &
\textbf{arXiv} &
\textbf{viXra} 
\\\hline

Avg. \# authors & 1.9 & 1.1
\\

Papers with one author & 35\% & 90\%
\\

University affiliation & 100\% & 35\%
\\

Collaboration across institutions & 35\% & 10\%
\\\hline
\end{tabular}
\vspace{2mm}
\caption{Authors, their affiliation and collaboration.}
\label{tbl:Area1}
\end{table}

\subsection {Publishing and Citations}\label{sec:PC}

Two-thirds of the arXiv papers have been published in journals. (See Table~\ref{tbl:Area2}.) On average they have received 1.3 citations. In contrast, very few viXra papers have been published in journals and have garnered few citations as a result. Instead of journal publishing, half of the viXra authors chose to self-publish on web pages or in web journals\footnote{viXra papers appearing in web journals are classified as Published (Web page or equivalent).}. None of the arXiv papers have done this\footnote{Making a PDF available is not considered web publishing for this metric.}.

\begin{table}[t!]
\centering
\begin{tabular}{m{5cm}m{2cm}m{2cm}}
\hline
\textbf{Metric} &
\textbf{arXiv} &
\textbf{viXra} 
\\\hline

Published (Journal or equivalent) & 65\% & 15\%
\\

Published (Web page or equivalent) & 0\% & 55\%
\\

Avg. \# of citations received & 1.3 & 0.11
\\

Has one or more citations & 35\% & 10\%
\\\hline

\end{tabular}
\vspace{2mm}
\caption{Publishing and citation (Google Scholar, May 1, 2012)}
\label{tbl:Area2}
\end{table}

\subsection {Writing Metrics}\label{sec:WM}
 
Many of the writing metrics were very similar between the two archives:  the average number of figures (per paper), numbered equations (per paper) and versions a paper had gone through. (Summarized in Table~\ref{tbl:Area3}.) arXiv papers were about 45\% longer than viXra and almost twice as likely to contain figures or tables. The largest difference was in the use of theorems and lemmas. Four arXiv papers contained a total of 30 while only one viXra paper contained one.

\begin{table}[t!]
\centering
\begin{tabular}{m{5cm}m{2cm}m{2cm}}
\hline
\textbf{Metric} &
\textbf{arXiv} &
\textbf{viXra} 
\\\hline

Avg. \# of versions of paper & 1.35 & 1.55
\\

Avg. \# of pages & 14.3 & 9.9
\\

Avg. \# of figures and tables & 3.5 & 3.9
\\

Has figures and tables & 80\% & 45\%
\\

Avg. \# of theorems and lemmas & 1.5 & 0.05
\\

Has theorems or lemmas & 20\% & 5\%
\\

Avg. \# of numbered equations & 24 & 28
\\

Has numbered equations & 90\% & 75\%
\\\hline

\end{tabular}
\vspace{2mm}
\caption{Writing metrics.}
\label{tbl:Area3}
\end{table}

\subsection {Referencing}\label{sec:Ref}

arXiv papers had twice as many references, pointed\footnote{For the purposes of this paper,~\cite{HW01} is an example of a pointer to a reference.} to them twice as often and self-referenced twice as often. (Table~\ref{tbl:Area4}) When averaged per page, the arXiv reference numbers were about 50\% larger than viXra's.

We found a number of citing behaviours in viXra papers not occurring in arXiv papers: all references were self references, the use of inline citations not in the references section\footnote{(Stephen Hawking,  A Brief History of Time, 1988) is an example of an inline citation not in the references section.} and the use of hyperlinks outside the references section. These occurred in a minority of papers, between one-sixth to one-third.

\begin{table}[t!]
\centering
\begin{tabular}{m{5cm}m{2cm}m{2cm}}
\hline
\textbf{Metric} &
\textbf{arXiv} &
\textbf{viXra} 
\\\hline

Avg \# of references & 26 & 11.5
\\

Has references & 100\% & 90\%
\\

Avg \# of inline pointers to\newline references & 37.7 & 20.8
\\

Has inline pointers to\newline references & 100\% & 85\%
\\

Has unused references & 0\% & 10\%
\\

Avg \# of self-references & 3.6 & 1.9
\\

Has self-references & 80\% & 55\%
\\

All references are self-references & 0\% & 15\%
\\

Total \# of inline citations not in\newline references & 0 & 21
\\

Papers with inline citations not\newline in references & 0\% & 25\%
\\

Total \# of hyperlinks outside\newline references section & 0 & 71
\\

Papers with hyperlinks outside\newline references section & 0\% & 30\%
\\\hline
\end{tabular}
\vspace{2mm}
\caption{Referencing.}
\label{tbl:Area4}
\end{table}

\subsection {Metrics Identifying Source Archive}\label{sec:MISA}

Metrics with large differences were extracted from Tables~\ref{tbl:Area1} to~\ref{tbl:Area4} to create the list of
indicators for the source archive. (Summarized in Table~\ref{tbl:RQ1}.)  The strong indicators for arXiv are unsurprising: university or equivalent affiliation and publication in a journal.  That viXra has low university affiliation, low journal publication rates and high self-publication rates, along with other stylistic differences from the other viXra-only indicators, indicate it has a more diverse pool of authors.  

\begin{table}[t!]
\centering
\begin{tabular}{m{5cm}m{2cm}m{2cm}}
\hline
\textbf{Metric} &
\textbf{arXiv} &
\textbf{viXra} 
\\\hline

University affiliation & 100\% & 35\%
\\

Published (Journal or equivalent) & 65\% & 15\%
\\

Has theorems or lemmas & 20\% & 5\%
\\

Collaboration across institutions & 35\% & 10\%
\\

Has one or more citations & 35\% & 10\%
\\

Papers with one author & 35\% & 90\%
\\

Published (Web page or equivalent) & 0\% & 55\%
\\

Papers with inline citations not\newline in references & 0\% & 25\%
\\

Papers with hyperlinks outside\newline references section & 0\% & 30\%
\\\hline
\end{tabular}
\vspace{2mm}
\caption{Metrics most likely to identify which archive a paper comes from.}
\label{tbl:RQ1}
\end{table}
 
\section{Related Work}\label{sec:RelatedWork}

arXiv has appeared in numerous studies.  In~\cite{GBMB09} the High Energy Physics (HEP) sub-categories were studied to determine the relative advantages of publishing in repositories versus open access journals. They concluded there was an \textit{``immense citation advantage''} for HEP papers in repositories.

One years worth of HEP papers from arXiv were studied in~\cite{MDVY06} to determine the share of HEP production by country and institution.

arXiv sends out daily email announcements of new papers.  The effect on long term citation count based on position within the announcement list was studied in~\cite{HG09,HG10}. They found there was an enhancement to the citation rate for papers at the beginning and end of the list.

We searched for, but could not find any papers studying viXra.

A scienceographic study of arXiv was carried out in~\cite{CMY12}. Latex files from 65,000 papers from the mathematics and computer science disciplines were analysed.  Items such as comments, authors and diagrams were quantified and compared between the two disciplines.  Other metrics such as the number of pages and number of Latex packages used were tracked over a 15 year period.

This work is very complimentary to~\cite{CMY12}.  A subset of metrics from each paper are the same (Avg. 
\# of authors/pages/theorems, \ldots). Where~\cite{CMY12} emphasizes Latex analysis (Comments, word counts, packages, \ldots) this work considers PDF-level analysis (References, citations, equations, \ldots). Where the metrics are the same, Table~\ref{tbl:Synthesis} synthesizes (with caveats) the results of the two papers.

\begin{table}[t!]
\centering
\begin{tabular}{m{5cm}m{2cm}m{2cm}m{3cm}m{2cm}}
\hline
\textbf{Metric} &
\textbf{arXiv\newline
Physics} &
\textbf{viXra} &
\textbf{arXiv\newline Mathematics} &
\textbf{arXiv\newline Computer Science}
\\\hline

Avg. \# authors & 1.9 & 1.1 & 1.2 & 1.7
\\

Percentage of papers with a\newline single author & 35 & 90 & More than half & 38
\\

Avg. \# of pages & 14.3 & 9.9 & 15 & 9
\\

Avg. \# of theorems and lemmas in papers containing theorems and lemmas & 7.5 & 1 & 5.5 & 4.9
\\

Percentage of papers with\newline theorems or lemmas & 20 & 5 & 71 & 48
\\\hline

\end{tabular}
\vspace{2mm}
\caption{Synthesis of compatible data. Columns \textit{arXiv: Mathematics} and \textit{arXiv: Computer Science} are from~\cite{CMY12}.Columns 2 and 3 were calculated from 20 papers each. Columns 4 and 5 were calculated from approx. 39,000 and 26,000 papers respectively.}
\label{tbl:Synthesis}
\end{table}

\newpage
\section{Threats to Validity}\label{sec:ThreatsToValidity}
 
After posing and answering the research question we were then responsible for considering threats to the validity of our results:

\textbf{Internal threats:} Any bias in experimental design could be an internal threat. One source of bias is selecting metrics favouring one archive over the other. This can be seen in extremely high or low numbers appearing consistently in results for arXiv or viXra. arXiv scored higher on 9 out of 14 criteria suggesting a bias in its favour.  Given this it is interesting to observe the four scores of zero, \{Publilshed on the Web or equivalent, number of inline references not in end references, all references being self-references and number of web links outside the references section\} are for arXiv. These categories favoured viXra. 

Papers selected may not be a representative sample of the physics papers in the arXiv and viXra archives.  To mitigate this we used a consistent selection process (matching sub-categories across archives, selecting year and month randomly from the last six (four) years and selecting the paper at numerical position 10) to minimize human bias.

\textbf{External threats:} Any bias hurting the generalizability of the results is an external threat.  arXiv scored 0 in 4 metrics. (See internal threats.) These metrics only appear in papers from the viXra archive so we cannot expect them to generalize to all archives.
 
 \newpage
\section{Conclusions}\label{sec:Conclusions}

This paper performed a scienceographic analysis of 20 papers from the physics category of the arXiv and viXra e-print archives.  Differences were found between the writing styles and contents of the papers of each. These differences are captured as a series of indicators. Many of the indicators are weak, appearing in one-third or less of papers. 

This paper makes two contributions: a first study of papers from the viXra archive and a scienceographic comparison of the papers from viXra and arXiv. 

\subsection{Future Work}

viXra's open access policies have attracted a population of non-academically trained authors~\cite{MW11}.  It would be interesting to verify this and determine if there are metrics indicating which group a paper falls into\footnote{Simply looking to see if the author is academically affiliated may not be enough.} \footnote{arXiv papers with their arXiv:NNNN.NNNNvN stamp in them have also been posted on viXra.}.  For example, are the 4 metrics where arXiv\footnote{Based on this work we are implicitly assuming all arXiv authors are academically trained.  Perhaps this isn't true.} papers scored 0 indicative of non-academically trained authors?  Creating a classifier using machine learning or evolutionary techniques could help answer this question.

Expanding the study to include more sub-disciplines from arXiv, viXra and other repositories could give further interesting insights into how science is written by different communities and sub-communities.

\section{Acknowledgements}

We thank Margaret Wertheim for providing the inspiration~\cite{MW11} to write this paper.


\bibliographystyle{alpha} 
\bibliography{bibliography}

\label{body_end}
\end{document}